\documentclass[doublecol]{epl2} 

\newcommand{\pii}{{\partial_i}}

\newcommand{\be}{\begin{equation}}
\newcommand{\ee}{\end{equation}}
\newcommand{\bea}{\begin{eqnarray}}
\newcommand{\eea}{\end{eqnarray}}

\usepackage{amsmath}
\usepackage{amsfonts}
\usepackage{amssymb}
\usepackage{bm}
\usepackage{color}
\usepackage{graphicx}

\title{Reactive Rayleigh-Taylor systems:  flame propagation and non-stationarity}

\author{A. Scagliarini\inst{1}, L. Biferale\inst{2,6}, F. Mantovani\inst{3}, M.
Sbragaglia \inst{2}, F. Toschi \inst{4,6} and R. Tripiccione \inst{5}}
\shortauthor{A. Scagliarini \etal}

\institute{
  \inst{1} Dept. of Physics and INFN, University of Tor Vergata, Via della Ricerca Scientifica 1, 00133 Rome, Italy and Dept. of Fundamental Physics, University of Barcelona, Carrer de Mart\'i i Franqu\`es 1, 08028 Barcelona, Spain \\
  \inst{2} Dept. of Physics and INFN, University of Tor Vergata, Via della Ricerca Scientifica 1, 00133 Rome, Italy \\
  \inst{3} Deutsches Elektronen Synchrotron (DESY), 
  D-15738 Zeuthen, Germany \\
  \inst{4} Dept. of Applied Physics and Dept. of Mathematics and
  Computer Science, Eindhoven University of Technology,
  5600 MB Eindhoven, The Netherlands  and CNR-IAC, Via dei Taurini 19, 00185 Rome, Italy\\
  \inst{5} Dept. of Physics and INFN, University of Ferrara, 
  I-44100 Ferrara, Italy \\
  \inst{6} International Collaboration for Turbulence Research (ICTR)

}


\abstract{Reactive Rayleigh--Taylor systems are characterized by the
  competition between the growth of the instability and the rate of
  reaction between cold (heavy) and hot (light) phases.  We present
  results from state-of-the-art numerical simulations performed at
  high resolution in 2d by means of a self consistent lattice
  Boltzmann method which evolves the coupled momentum and thermal
  equations and includes a reactive term.  We tune the
  parameters affecting flame properties, in order to address the competition
  between turbulent mixing and reaction, ranging from slow to
  fast-reaction rates. We also study the mutual feedback between
  turbulence evolution driven by the Rayleigh-Taylor instability and
  front propagation against gravitational acceleration. We
  quantify both the enhancement of flame propagation due to turbulent
  mixing for the case of slow reaction-rate as well as the slowing down
  of turbulence growth for the fast reaction case, when the flame
  quickly burns the gravitationally unstable phase. An increase of
  intermittency at small scales for temperature characterizes the case of
  fast reaction, associated to the formation of sharp wrinkled
  fronts separating pure hot/cold fluids regions.}

\begin{document}

\maketitle

\section{Introduction}
Many natural and industrial processes involve fluid transport and
mixing of passive or active scalar fields; examples include
concentration fields of chemicals or biological species as well as the
temperature field in natural convection. While many of these phenomena
have been the subject of in-depth studies, the cases where chemical
reactions are involved, presenting an even richer phenomenology, have
received considerably less attention.  We address the problem of the
interplay of reaction and turbulent mixing in Rayleigh-Taylor (RT)
systems (a situation occurring, for example, in thermonuclear burning
of type Ia supernovae \cite{zingale,khokhlov,gamezo} or in the
inertially confined nuclear fusion \cite{freeman}) focusing
on the different regimes which develop as we vary the ratio between
the characteric time scales of underlying turbulence, $\tau_{turb}$,
and the reaction time, $\tau_R$.  We limit this study to the case of
single-step reaction, i.e. the two reactant scalar fields are
distinguished by a reaction progress variable, proportional to the
temperature (see Fig. \ref{fig:RTflames}).  The two different
temperatures in the hot and cold blobs of fluid of our numerical
setup, mimick the combustion of a cold mixture of actual reactants
into a hot mixture of burnt products
\cite{cetegen,tieszen,chertkovflames}.  The interesting point in this
set-up is given by the natural competition between gravitational
forces, which tends to mix the fluid and to produce a larger and
larger mixing layer with uniform temperature, and combustion, which
works against this mixing, trying to burn the whole volume and
producing a propagating flame of given tickness and velocity.
Moreover, the global phenomenology is complicated by the natural
unsteadiness of the underlying RT problem.  The
Damk$\ddot{\mbox{o}}$hler number, $Da$, is the natural control
parameter and is identified by the ratio between the turbulent time
scale, $\tau_{turb}$ and the reaction time scale $\tau_R$.  Notice that
because of RT unsteadyness the $Da$ number depends on time:
$$
Da(t) =\tau_{turb}(t)/\tau_R.
$$ 
where $\tau_{turb} \propto t $, as of standard RT phenomenology
\cite{chertkovflames}.

We perform highly resolved numerical
simulations in 2d, with a resolution up to $4096 \times 10000$
grid points
(see table \ref{table:param}).
\begin{table*}
  \begin{center}
    \begin{tabular}{|c | c c c c c c c c c |}
      \hline & $At$ & $L_x$ & $L_z$ & $\nu$ & $g$ & $T_u$ & $T_d$ & $\tau_R$ & $ \tau $ \\
      \hline run A & $0.05$ & 4096 & 10000 & 0.005 & $2.67 \times 10^{-5}$ & $0.95$ & $1.05$ & $5\times 10^3$ &
      $5.5 \times 10^4$ \\
      run B &$0.05$& 4096 & 10000 &0.005 & $2.67 \times 10^{-5} $ &
      $0.95$ & $1.05$ & $5 \times 10^4 $ & $ 5.5 \times 10^3$ \\ 
      run C &$0.05$& 4096 & 10000 &0.005 & $2.67 \times 10^{-5} $ &
      $0.95$ & $1.05$ & $5 \times 10^5 $ & $ 5.5 \times 10^3$ \\ \hline
    \end{tabular}
    \caption{Parameters for the three sets of runs. Atwood number,
      $At=(T_d-T_u)/(T_d+T_u)$; viscosity $\nu$ (thermal diffusivities
      are the same since Prandtl number is $1$ for each run; gravity
      $g$; temperature in the upper half region, $T_u$; temperature in
      the lower half region, $T_d$; reaction characteristic time
      $\tau_R$; normalization time, $\tau=\sqrt{L_x/(g\;At)}$.}
    \label{table:param}
  \end{center}
\end{table*}
The 2d set up allowed us to reach a wide scale separation and a
time-span large enough to address problems at both small
and large Damk$\ddot{\mbox{o}}$hler numbers, something still unfeseable in
3d.  Our study has also a methodological motivation. We adopted a {\it
  fully consistent} thermal lattice Boltzmann method to evolve
simultaneously the momentum equations and the
advection-diffusion-reaction equations for temperature. The novelty
here is to show that the method works well also in a non trivial
case where the thermal modes are directly forced by the combustion
terms.

The main result of the paper concerns the quantification of the front
propagation due to turbulence enhancement for the slow reaction case,
$Da \ll 1$, and the clear signature of a strong feedback on the fluid
evolution induced by the flame propagation when $Da > 1
$. In the latter case, we also measure an important increase of the temperature
intermittency at small scales.

\section{Equations of motion and numerical setup}
We adopt a numerical scheme based on a recently proposed thermal
lattice Boltzmann algorithm \cite{JFM,POF}, which is able to reproduce
the correct thermohydrodynamics of an ideal gas with good numerical
accuracy \cite{POF2}.  To do that, the probability densities $f_l({\bm x},t)$
for a particle with velocity ${\bm c}_l$ (belonging to
a discrete set, with the index $l$ running over 37 values \cite{POF})
at space location ${\bm x}$ and  time $t$ evolve according to
the lattice Boltzmann BGK equation \cite{BGK,wolf,succi}
\be\label{eq:LBGK} 
f_l({\bm x}+{\bm c}_l \Delta t, t+\Delta t) - f_l({\bm x},t) = -\frac{\Delta t}
{\tau_{LB}}\left(f_l - f_l^{(eq)} \right)({\bm x},t);
\ee
the lhs stands for the free streaming of particles and the rhs
represents the relaxation process towards  Maxwell equilibrium
$f_l^{(eq)}({\bm x},t)$ with a characteristic time $\tau_{LB}$
($\Delta t$ is the simulation time step).  Once the density ($\rho$),
velocity (${\bm u}$) and temperature ($T$) fields are defined in terms
of the lattice Boltzmann populations as 
\be 
\rho = \sum_l f_l; \; \rho{\bm u} = \sum_l f_l {\bm c}_l; \; D\rho T = \sum_l f_l\left|{\bm c}_l - {\bm u} \right|^2, 
\ee 
($D$ is the number of space dimensions), it has
been shown \cite{JFM,POF} that the following set of macroscopic
equations can be recovered (repeated indexes are summed upon): 
\be \label{eq:eqTHY}
\begin{cases}
  D_t \rho = - \rho \pii u_i \\
  \rho D_t u_i = - \pii p - \rho g\delta_{i,z} + \nu \Delta u_i  \\
  \rho c_v D_t T  = k \Delta T + \frac{1}{\tau_R} R(T) ,
\end{cases}
\ee 
where $k$ is the thermal conductivity, $\tau_R$ the reaction time, $g$ the
gravity and $c_v=D/2$ the specific heat at constant volume. Eqs (\ref{eq:eqTHY})
are obtained provided that the thermohydrodynamic fields appearing in the
equilibrium density functions are properly shifted:
\be 
{\bm u } \rightarrow {\bm u} + \tau_{LB} {\bm g}; \; T \rightarrow T + \frac{\tau_{LB} (\Delta t - \tau_{LB})}{D} g^2 + \frac{\tau_{LB}}{\tau_R} R(T).  
\ee 
The novelty here is in the extra reaction term introduced for the temperature
field 
\footnote{This shift represents  a kind of implicit equation, since, in
principle, $R(T)$ should be a function of the ``real'' thermodynamic
temperature, which must be shifted itself \cite{POF}; however, we can overcome
this problem, observing that for $\tau_R \gg \tau_{LB}$ (always true in actual
situations), and since the other shift is $\mathcal{O}(g^2)$  (hence
$\mathcal{O}(Kn^2)$ \cite{POF}) we can safely assume that $R(T^{(H)}) \approx
R(T)$. }.  

In the third equation of (\ref{eq:eqTHY}) we have already subtracted the
compression term $p \nabla \cdot {\bm u}$ to avoid effects due to a varying heat
capacity or global heating of the system coming from steady increase of the
underlying mean pressure.

\begin{figure}
\begin{center}
  \advance\leftskip-0.55cm
  \onefigure[scale=0.6]{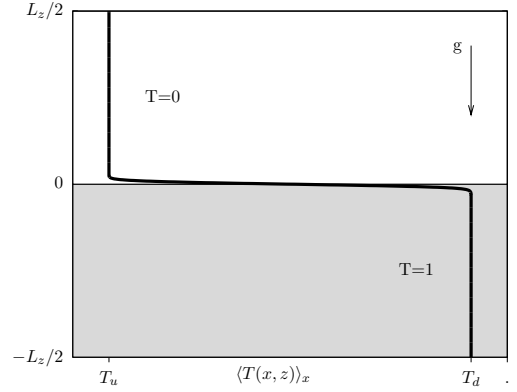}
  \caption{Initial configuration for the Rayleigh-Taylor system with
    combustion: cold fluid (fresh fuel) at $T=0$ on top and hot fluid
    (burnt material) at $T=1$ on bottom. Such temperature jump at the
    interface is smoothed by a hyperbolic tangent profile with a width
    of the order of 10 grid points and with a randomly perturbed
    centre (thus enabling to perform independent runs). The system used is two dimensional and has size $L_x \times L_z$ plus periodic boundary conditions 
    applied in the streamwise ($x$) direction. The fluid used is an ideal gas.}
\label{fig:RTflames}
\end{center}
\end{figure}

The reaction rate must be zero in the pure phases, which we set at temperatures
$T=1$, for the hot fluid at the bottom, and $T=0$, for the cold fluid on top
(see figure \ref{fig:RTflames}), so that $R(0)=R(1)=0$; it must also transforms,
irreversibly, the pure cold phase (unstable) into the hot one (stable).
A simple model for $R(T)$ with these properties is given by a
logistic-type expression $$R(T) = T(1-T)$$
originally proposed \cite{fisher,KPP} as a model for the propagation
of an advantageous gene in a population. 

We performed three high resolution sets of runs (on lattices of $4096 \times
10000$ grid points) on the QPACE Supercomputer \cite{QPACE1,LBMonQPACE}, with
different reaction times (run parameters are collected in table
\ref{table:param}).  For each set, we carried out several ($\mathcal{O}(10)$)
independent runs, in order to enhance statistics.
 
\section{Results and discussion}
\label{sec:flamesresults} 
Any RT system, even in the case of $\tau_R \gg 1$, will eventually reach the
fast reaction limit, i.e. a situation where $Da(t) \gg 1$. This is due to the
fact that the underlying turbulence slows down adiabatically, $\tau(t) \propto
t$.  As a consequence, sooner or later the flame tends to become active, burning
at a rate faster than the turbulence stirring/mixing. Here we study the two
regimes $Da \ll 1$ and $Da > 1$ and the transition between them.

\subsection{Mean temperature profiles evolution}
\begin{figure}
  \begin{center}
    \onefigure[scale=0.5]{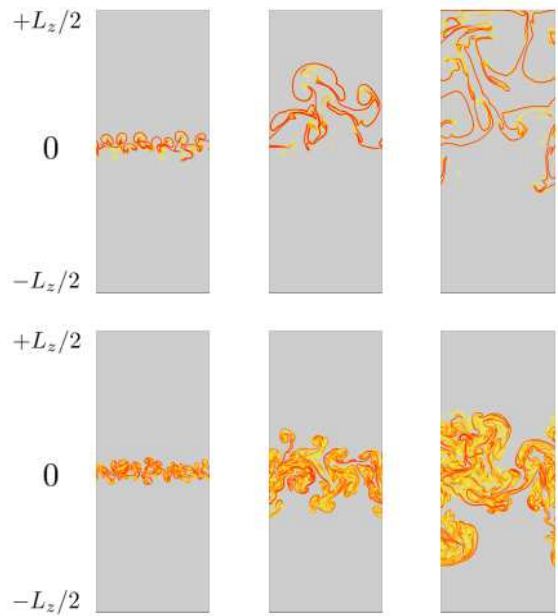}
    \caption{Snapshots of the magnitude field of the temperature
      gradient for the fast reaction case, run A (top panel), and for
      the non--reacting case (bottom panel).}
    \label{fig:gradT}
  \end{center}
\end{figure}

\begin{figure}
  \begin{center}
    \advance\leftskip-0.55cm
    \onefigure[scale=0.5]{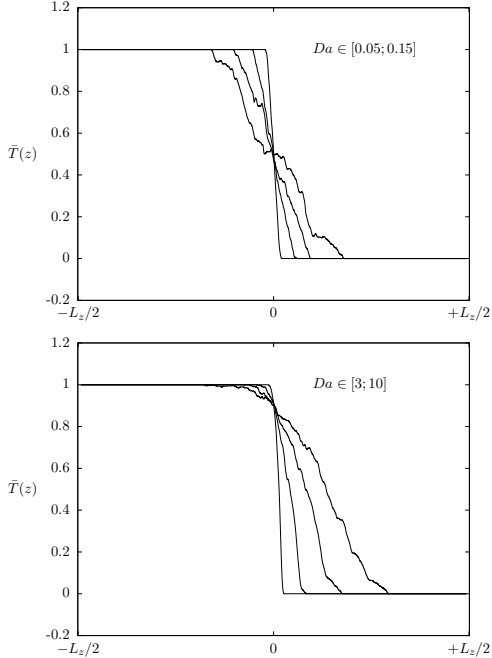}
    \caption{Mean temperature profiles at various times for run A (bottom
    panel) and run C (top panel). The latter case is almost identical to the
    non-reacting case.}
    \label{fig:profiles}
  \end{center}
\end{figure}

\begin{figure}
  \begin{center}
    \advance\leftskip-0.55cm
    \onefigure[scale=0.6]{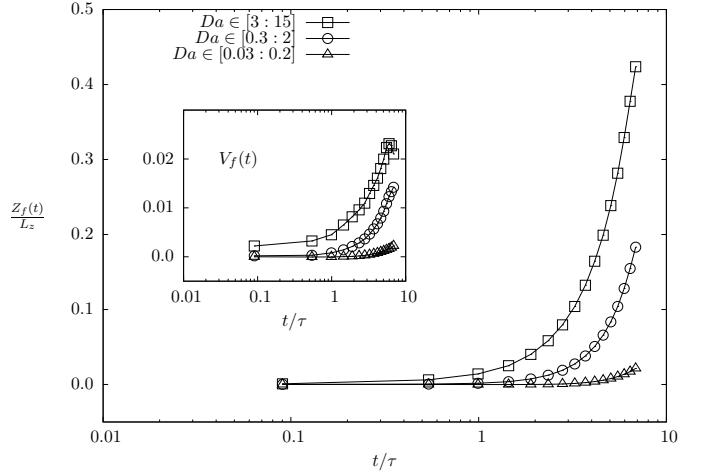}
    \caption{Reaction front coordinate $Z_f$ (normalized by the total vertical
      box length) and front speed $V_f(t)$ (inset) as a function of time for the
      three sets of runs: the faster the reaction, the more rapidly $Z_f$ and
      $V_f$ grow in time. }
    \label{fig:zfront}
  \end{center}
\end{figure}

For large $Da$, the mixing is effective only at very small scales (where the
characteristic times of the fluid motion are shorter), while the reaction tends
to make uniform the mixed regions: as a result we get a topology of the
temperature field which is made of ``patches'' separated by rather thin
interfaces, which are smoother than the non-reacting RT case
\cite{chertkovflames}; in addition, the front of the hot phase moves, on
average, with a non zero mean drift velocity towards the top. These preliminary
features can be better understood, at a pictorial level, looking at figure
\ref{fig:gradT}, where we show the magnitude of the temperature gradient $
\left| \nabla T \right|^2 = ((\partial_x T)^2 + (\partial_z T)^2)$ at three
different times in the evolution for the fastest reaction rate that we have
studied (top panel), and compare it with the non-reacting case (bottom panel).
On the other hand, the larger the reaction time $\tau_R$ the closer is
the phenomenology to the standard RT case: to see this we compare in
figure \ref{fig:profiles} the evolution of the mean temperature
profile 
\be 
\bar{T} (z,t) = \frac{1}{L_x} \int T(x,z;t)dx 
\ee 
for the two extreme cases in our database, runs A and C: while for $\tau_R = 5
\times 10^5$ the evolution is basically undistinguishable from the usual RT
dynamics\cite{POF,POF2}, in the fast rection case ($\tau_R = 5 \times 10^3$) the
center of mass of the system clearly moves upwards, due to the burning
processes, causing a shift --and an asymmetry-- of the mixing region.
The propagation of the burnt hot material {\it front} against the fresh reactant
($T=0$) can be quantified by the barycentric coordinate $Z_f(t)$, that we define
as the following integral \cite{koudella,constantin}: 
\be  \label{eq:zfront} 
Z_f(t) =
\int_{-L_z/2}^{+L_z/2} \bar{T}(z,t)dz 
\ee 
in figure \ref{fig:zfront} we plot the function $Z_f(t) \; \mbox{vs} \; t$ for the three different reaction rates: the growth of $Z_f(t)$ in time is greatly
enhanced when going from small to large $\tau_R$.

\subsection{Front propagation speed}
\begin{figure}
  \begin{center}
    \advance\leftskip-0.55cm
    \onefigure[scale=0.7]{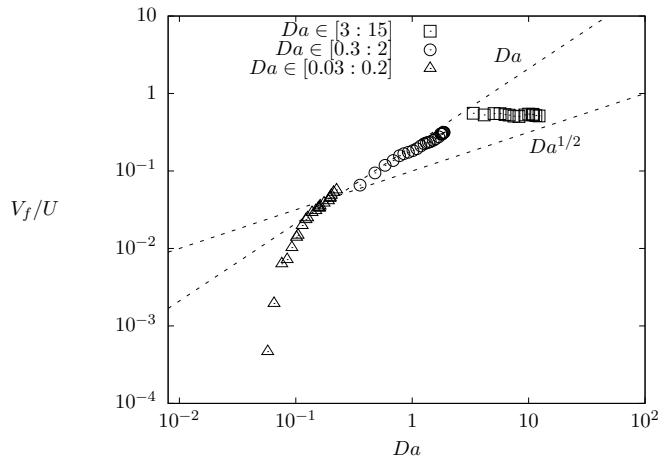}
    \caption{Front speed normalized by the root mean square (vertical)
      fluid velocity for the three runs as a function of the
      Damk$\ddot{\mbox{o}}$hler number $Da(t)$.  The solid line
      represents the theoretically predicted behaviour $V_f/U \propto
      Da$, obtained on the basis that for $Da <1$ flame propagates
      inside the well mixed mixing layer. The prediction $V_f/U
      \propto Da^{1/2}$, obtained from the assumption that in the
      pre-mixed combustion (slow reaction) regime one can simply
      substitute the molecular diffusivity with the turbulent one
      $\kappa \rightarrow \kappa_T$ in the expression for the laminar
      flame speed, is also plotted (dashed line).}
    \label{fig:VvsDa}
  \end{center}
\end{figure}

If we integrate eqs. (\ref{eq:eqTHY}) over the whole volume, and
divide by $L_x$, we get an exact equation for the propagating front
speed:
\begin{eqnarray} \label{eq:frontspeed1} V_f(t) = \partial_t \left(
    \int_{-L_z/2}^{+L_z/2} \bar{T}(z,t)dz\right) = \frac{1}{\tau_R}
  \langle {T(1-T)} \rangle,
\end{eqnarray}
(where $\langle (\cdot) \rangle = (1/L_x)\int\int (\cdot) dxdz$) since
the boundary terms vanish, owing to the periodic conditions on the
lateral walls and to the adiabatic condition at top and bottom plates
($\nabla T |_{z=\pm L_z/2} = 0$).  For the laminar flame (that is
without gravity producing turbulence), the integral can be evaluated
exactly (using, for instance, the usual hyperbolic tangent profile) to
give an explicit expression for the speed, that is $V_f \propto L_f/\tau_R$,
where $L_f$ is the flame thickness: as the latter can be estimated to
be $L_f \propto \sqrt{\kappa \tau_R}$, we end up with the well known result:
\be \label{eq:frontspeed3} 
V_f \propto \sqrt{\frac{\kappa}{\tau_R}}, 
\ee
that is the flame propagates at constant speed. \\ We now ask what changes when
turbulence sets is.  In the small $Da$ limit, when turbulence has the time to
mix the fluids before reaction becomes active, we are in the so-called {\it
pre-mixed combustion}. In this case, it has been conjectured
\cite{koudella,damkohler} that the simplest way to extend the result of the
laminar case is to replace in expression (\ref{eq:frontspeed3}) the molecular
diffusivity $\kappa$ with an effective (turbulent) eddy diffusivity $\kappa_T$.
If we use for the latter the dimensional estimate $\kappa_T(t) \sim U(t) L(t)$,
where $U$ and $L$ are large scale characteristic velocity and length (in our
case, e.g. the root mean square velocity and the mixing region length), and plug
it into (\ref{eq:frontspeed3}), we get: 
\be 
V_f(t)
\sim \sqrt{\frac{\kappa_T}{\tau_R}} \sim
\sqrt{\frac{U(t)L(t)}{\tau_R}} \sim U(t)
\sqrt{\frac{(L(t)/U(t))}{\tau_R}}, \ee where \be V_f(t) \sim U(t)
\sqrt{\frac{\tau_{turb}}{\tau_R}} \equiv U(t) \, Da(t)^{1/2}.  
\ee
This prediction, probably valid to describe the evolution of slow
flames in stationary turbulent flows is unlikely to be relevant for RT
turbulence. The reason is that in order to observe a ``eddy-diffusivity'' driven flames one needs also 
scale separation between the turbulent eddies  and the flame tickness, something that is not realized by 
the evolving RT system.
On the other hand, 
 we can rewrite (\ref{eq:frontspeed1})
exactly as: 
\be 
\label{eq:frontspeed2.a} V_f(t) = \frac{1}{\tau_R}
[\langle {\bar T (1- \bar T)} \rangle - \langle \overline{\theta^2}
\rangle ].  
\ee 
where with $\theta = T -\bar T$ we denote the fluctuations with respect to the
mean vertical profile. It is clear now that for $Da <1$, the flame cannot have
any strong influence on the underlying RT evolution and we can identify the
first term on the rhs as the mixing layer length $L(t) = \langle {\bar T (1-
\bar T)} \rangle$. Moreover, we know that in RT temperature fluctuations are
almost constant in time and homogeneous inside the mixing layer
\cite{chertkovprl}, so also the second term on the rhs is proportional to the
mixing layer extension. A natural prediction for $Da <1$ is therefore:
\begin{equation}
\label{eq:new}
V_f(t) \propto \frac{L(t)}{\tau_R}; \qquad V_f(t)\propto U(t) Da(t)
\end{equation}
In standard (stationary) turbulent reacting systems, one can check this
prediction against experiments/simulations at various $Da$, obtained changing
the reaction rate or the underlying turbulent intensity, while in our reacting
RT setup we can exploit the fact that $Da=Da(t)$ varies in time. In figure
\ref{fig:VvsDa} we plot the front speed (normalized with the root mean square
velocity) as a function of $Da$ (which is itself a function of the
simulation time) for the three runs.
As one can see, our prediction (\ref{eq:new}) works satisfactorily 
in a wide range of
$Da(t)$, showing deviations only for very small times, where turbulence is not
yet developed and the flame evolution is strongly influenced by the initial
configuration, and for $Da(t) >1$ where it cannot be expected to be valid.  
In the latter case our data point flatten, as
we clearly observe the feedback of the flame on the turbulent
evolution, with a sort of synchronization between flame propagation and
evolution of the turbulent kinetic energy toward a value where $V_f(t) \sim
U(t)$. Such a behaviour turns out to be in agreement with recent theoretical results 
obtained through a mean-field approach \cite{brandenburg}.

\subsection{Small scale intermittency}
When the reaction rate is fast ($Da \gg 1$), there are no extended regions
which are well mixed, since the cold material, as soon as it is
slightly entrained through the hot one is suddenly burnt. As a result,
the temperature field organizes in patches of pure reactants/products
separated by sharp interfaces (being in the so called ``segregated
regime''), and, consequently, it has been conjectured that an increased
intermittency develops at the small scales \cite{chertkovflames}. 
The authors in \cite{chertkovflames} also derived a phenomenological prediction
for the scaling laws of fluid temperature (and velocity) structure functions,
according to which, in the asymptotics of $Da \gg 1$, they should follow the
relation
\be 
\label{eq:sfflames} S^{(p)}_{T}(R,t) \equiv \langle |\delta_R T|^p
\rangle \sim \left(\frac{R}{L(t)} \right) ^{2/3}, 
\ee 
(where $L(t)$ is the mixing length), irrespective of the order $p$. From
Eqn. (\ref{eq:sfflames}) the expression for the flatness reads:
\be 
\label{eq:flatnessflames} 
F_T^{(p)}(R,t) = \frac{\langle |\delta_R  T|^p \rangle}{\langle |\delta_R T|^{p/2} \rangle^2} \sim R^{-2/3}
L(t)^{2/3} 
\ee 
and so it increases with decreasing $R$ for all orders, a clear indication of
strong small scales intermittency.
\begin{figure}
  \begin{center}
    \advance\leftskip-0.55cm
    \onefigure[scale=0.7]{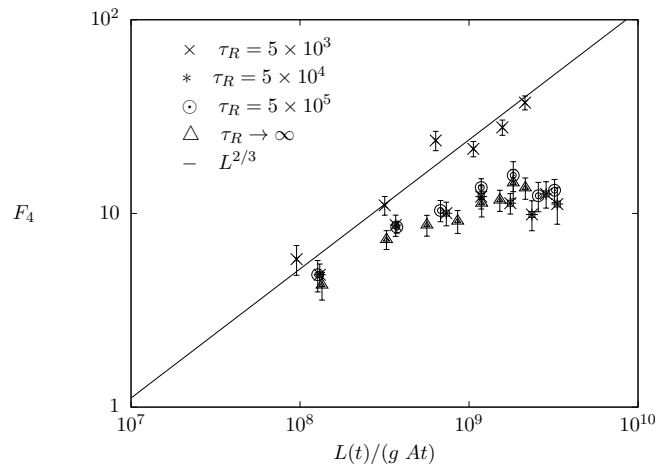}
    \caption{The $4$-th order flatness $F_4$ for the three runs and
      for the non-reacting RT ($\tau_R \rightarrow \infty$). Data from
      run A ($Da \gg 1$) agree  well, within error bars,
      with the prediction given by equation (\ref{eq:flatnessflames})
      $F_4 \sim L^{2/3}$.}
    \label{fig:flatnessflames}
  \end{center}
\end{figure}
In figure \ref{fig:flatnessflames} we show the growth of $F_4$ as function of
the mixing length $L$, for the three runs: the flatness for run A, corresponding
to the smallest reaction time, is in good agreement, within error bars, with the
prediction of equation (\ref{eq:flatnessflames}), $F_T \sim L^{2/3}$; instead,
at increasing $\tau_R$, intermittency is depleted and the flatness grows more
slowly, at a rate comparable (within error bars) with the non-reacting RT case,
whose data are also reported for comparison.

\section{Conclusions}
We used a self consistent thermal lattice Boltzmann algorithm to perform numerical simulations of 2D Rayleigh--Taylor turbulence, in presence of chemical
reactions between hot and cold fluids. The reaction was modelled by
means of a Fisher-Kolmogorov-Petrovsky-Piskunov source term in the
temperature equation; this term has been introduced by application of
a suitable shift of the temperature field appearing in the equilibria
of the lattice Boltzmann equation.

We analyzed the crossover among the various regimes emerging from the
competition of turbulent mixing and reaction, going from the segregated ($\tau_R
\gg \tau_{turb}$) to the well mixed one ($\tau_R \ll \tau_{turb}$). We showed
that, in the latter case, the effect of turbulence is to enhance the reaction
front speed leading to an homogeneous burning in the whole mixing layer region. 
On the other hand, for moderate and large Damk$\ddot{\mbox{o}}$hler, there is a
feedback of the reaction on the statistical properties of the temperature field,
resulting in increased intermittency at small scales in reasonable
accordance with the prediction of
\cite{chertkovflames}.

\section{Acknowledgments}
We  thank A. Celani and M. Cencini for useful suggestions and 
the QPACE development team for support during the implementation
of our code and execution of the runs. We acknowledge access to the QPACE and
eQPACE systems.

\end{document}